\documentclass[journal=jpcafh,manuscript=article]{achemso}
\setkeys{acs}{usetitle=true}

\usepackage{amsmath}
\usepackage{amssymb}
\usepackage{graphicx}
\usepackage{setspace}
\usepackage{color}
\usepackage[usenames,dvipsnames]{xcolor}

\usepackage{arydshln}

\DeclareFontFamily{OT1}{pzc}{}
\DeclareFontShape{OT1}{pzc}{m}{it}%
              {<-> s * [1.100] pzcmi7t}{}
\DeclareMathAlphabet{\mathpzc}{OT1}{pzc}%
                                 {m}{it}

\newcommand{\mx}[1]{\boldsymbol{#1}}

\newcommand{\mr}[1]{\mathrm{#1}}

\def\cm{cm$^{-1}$}
\def\Eh{E$_\mathrm{h}$}

\def\nperm{{N_\mathrm{perm}}}

\def\nbas{{N_\mathrm{b}}}
\def\npart{{n_\mathrm{p}}}
\def\som{Supporting Information}

\def\X1Sgp{X\ ^1\Sigma_\mr{g}^+}
\def\B1Sup{B\ ^1\Sigma_\mr{u}^+}
\def\b3Sup{b\ ^3\Sigma_\mr{u}^+}
\def\a3Sgp{a\ ^3\Sigma_\mr{g}^+}
\def\Sp{S_\mr{p}}
\def\Se{S_\mr{e}}

\def\colgray{\color{gray}}
\def\colblack {\color{black}}

\author{Edit M\'atyus}
\email{matyus@chem.elte.hu}
\affiliation[E\"otv\"os University]
{%
  Laboratory of Molecular Structure and Dynamics,
  Institute of Chemistry, E\"otv\"os University,
  P.O.~Box 32, H-1518, Budapest 112, Hungary
}

\title[Resonances in Pre-Born--Oppenheimer Theory]%
{On the Calculation of Resonances in Pre-Born--Oppenheimer Molecular
Structure Theory}

\begin{document}
\begin{abstract}
  The main motivation for this work is the exploration of rotational-vibrational states
  corresponding to electronic excitations in
  a pre-Born--Oppenheimer quantum theory of molecules.
  These states are often embedded in the continuum of the lower-lying dissociation channel of
  the same symmetry, and thus are thought to be resonances.
  In order to calculate rovibronic resonances,
  the pre-Born--Oppenheimer variational approach of [J. Chem. Phys. 137, 024104 (2012)],
  based on the usage of explicitly correlated Gaussian functions and the global vector representation,
  is extended with the complex coordinate rotation method.
  The developed computer program is used to calculate resonance energies and widths for
  the three-particle positronium anion, Ps$^-$, and the four-particle positronium molecule, Ps$_2$.
  Furthermore, the excited bound and resonance rovibronic states of the
  four-particle H$_2$ molecule are also considered.
  Resonance energies and widths are estimated for the lowest-energy resonances of H$_2$
  beyond the $\b3Sup$ continuum.

  \vspace{1cm}
\end{abstract}


%
%
\clearpage
\section{Introduction\label{ch:intro}}
The present work is devoted to conceptual and computational problems
in pre-Born--Oppenheimer (pre-BO) molecular structure theory.
Without the Born--Oppenheimer (BO) approximation,\cite{BoOp27,Born51,BoHu54}
in a ``pre-BO world'',
we can gain in accuracy for the numerical results but
loose a central paradigm for the well-accustomed concepts of chemistry.
The reconstruction or interpretation of many common chemical concepts becomes a real challenge.
One of these famous challenges, the problem of the quantum structure of molecules
has been recognized long ago\cite{Wo76,Wo77b,WoSu77,We84} and studied by many authors.\cite{ClDi80,Wo80,Wo86,
CaAd04,SuWoCPL05,UMH06,UMH08,Lu12,CaCh98,GoSh11,GoSh12,MaHuMuRe11a,MaHuMuRe11b}

\begin{figure}
  \includegraphics[width=0.8\linewidth]{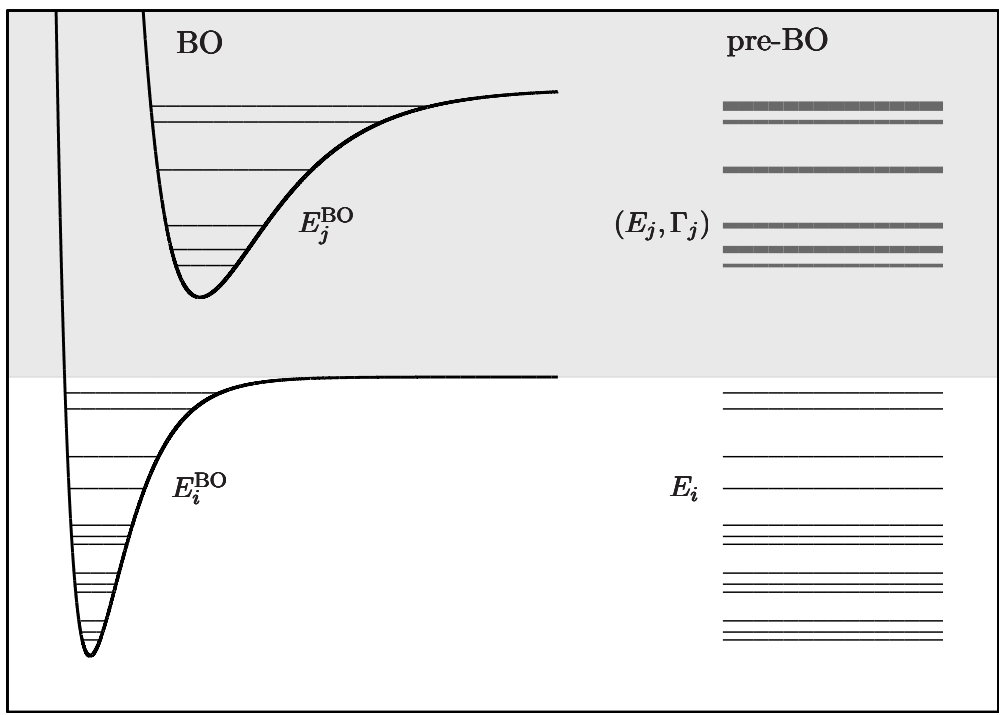}
  \caption{%
    Motivation for this work:
    calculation of rovibrational levels corresponding to
    electronically excited states, which are bound in the Born--Oppenheimer description
    but which appear as resonances in
    pre-Born--Oppenheimer molecular structure theory.
    \label{fig:preBOreson}
  }
\end{figure}

In the present work we address another challenge,
the status of electronically excited rotational-vibrational states
in a pre-BO quantum mechanical description.
In a pre-BO description there are no electronic states with corresponding
potential energy curves or surfaces on which the rotational-vibrational Schr\"odinger equation could be solved.
In addition, rotational-vibrational states corresponding to electronic excitations are often embedded
in the lowest-lying dissociation channel of the system, \ref{fig:preBOreson},
prone to predissociation.\cite{Harris63}
These rovibronic states are thus accessible in a pre-BO description only as resonances.
These rovibronic resonances are characterized
with some energy and width corresponding to a finite non-radiative (predissociative) lifetime.
Our aim is to explore how these properties can be calculated in a pre-BO approach.

As to the numerical results, there are practical approaches used for the calculation of quasi-bound states
in molecular science \cite{KuKrHo88,Rein82,Moi98}.
The stabilization technique, a very simple computational tool has been used to identify resonances
and to estimate the resonance energy\cite{HaTa70,Ho83,UsSu02} and 
can also be extended to the calculations of the width.\cite{MaRaTa93,MaTaRyMo94,MuYaBu94,RyMoMaTa94}
The complex coordinate rotation method \cite{AgCo71,BaCo71,Si72}
is a neat mathematical approach for the calculation
of the resonance energy and width, and has been used in several cases,\cite{Rein82,Moi98}
for example in rotational-vibrational
calculations on a potential energy surface within the Born--Oppenheimer (BO) theory.\cite{WaBo95,RyMoMaTa94}
The usage of complex absorbing potentials \cite{RoEnMo90,RoLiMo91,RiMe93}
has been a popular technique
in molecular spectroscopy and quantum reaction kinetics with many applications.\cite{SeMi91,ThMi97,Miller98}
There exist also more involved and specialized approaches, such as the  solution of
the Faddeev--Merkuriev integral equations. \cite{Papp02,Papp05}

The present work is organized as follows. First, the
necessary theoretical framework is described for the variational solution of
the Schr\"odinger equation for bound states of few-particle systems.
Then, this approach is extended for the identification and calculation of quasi-bound states.
Next, numerical examples are presented for the three-particle Ps$^-$ and the four-particle Ps$_2$.
Finally, the description of excited bound and resonance rovibronic states of the four-particle H$_2$
is explored.
In the end, we finish with a summary and outlook.

%
%
\clearpage
\section{Theory and computational strategy\label{ch:theo}}
The Schr\"odinger equation for an $(\npart+1)$-particle system with $m_i$ masses and $q_i$ electric charges
assigned to the particles is
\begin{align}
  \hat{H} \Psi = E \Psi
\end{align}
with the non-relativistic quantum Hamiltonian in Hartree atomic units
\begin{align}
  \hat{H}
  =
  \hat{T} + \hat{V}
  =
  -\sum_{i=1}^{\npart+1}\frac{1}{2m_i}\Delta_{\mx{r}_i}
  +\sum_{i=1}^{\npart+1}\sum_{j>i}^{\npart+1} \frac{q_iq_j}{|\mx{r}_i-\mx{r}_j|}
\end{align}
written in laboratory-fixed Cartesian coordinates $\mx{r}=(\mx{r}_1,\mx{r}_2,\ldots,\mx{r}_{\npart+1})$.

In the present work we use the bound-state variational approach of
Ref.~\citenum{MaRe12}
and
(a) combine it with the stabilization technique to quickly identify long-lived resonances;
and
(b) extend it with the complex coordinate rotation method to
calculate resonance energies and widths.

\subsection{Variational pre-BO calculations using explicitly correlated Gaussian functions and the global vector representation}
%
The overall translation of the center of mass is eliminated by writing the kinetic energy operator in
terms of Jacobi Cartesian coordinates and the translational kinetic energy of the center of mass
is subtracted.
As an alternative to this approach, the original laboratory-fixed Cartesian coordinates
can be used throughout the calculations without any further coordinate
transformation employing a special elimination technique for the overall translation during the evaluation of
the matrix elements.\cite{SiMaRe13}

The matrix representation of the translationally invariant Hamiltonian is constructed using
a symmetry-adapted basis set defined as follows.

A basis function with
the quantum numbers%
\footnote{%
  $N$ denotes here the total spatial (orbital plus rotational) angular momentum quantum number in
  agreement with the recommendations of the
  International Union of Pure and Applied Chemistry.\cite{GreenBook07} 
  We note that in Ref.~\citenum{MaRe12}
  the symbol $L$ was used in the same sense.
  Furthermore, $p$ is the parity, which we call ``natural'' if $p=(-1)^N$  and ``unnatural'' if $p=(-1)^{N+1}$.
  In this work we restrict the discussion to natural-parity states.
  The total spin quantum number for particles of type $a$ is $S_a$.
  For example, $S_\mr{p}$ and $S_\mr{e}$ denote the total spin quantum numbers
  for the protons and the electrons, respectively.
}
$\lambda=(N,M_N,p)$ and $\varsigma=(S_a,M_{S_a},S_b,M_{S_b},\ldots)$
($a,b,\ldots$ label the particle type) is constructed as
\begin{align}
  \Phi^{[\lambda,\varsigma]}(\mx{r},\mx{\sigma})
  =
  \hat{\mathcal{A}}
  \lbrace
    \phi^{[\lambda]}(\mx{r})\ \chi^{[\varsigma]}(\mx{\sigma})
  \rbrace \ ,
  \label{eq:basis}
\end{align}
where $\hat{\mathcal{A}} = (\nperm)^{-1/2} \sum_{i=1}^{\nperm}\varepsilon_i \hat{P}_i$
is the symmetrization and antisymmetrization operator
for bosonic and fermionic-type particles, respectively. $\hat{P}_i$ ($i=1,2,\ldots,\nperm$)
is an operator permuting identical particles and $\varepsilon_i=-1$ if $\hat{P}_i$ corresponds
to an odd number of interchanges of fermions, otherwise, $\varepsilon_i=+1$.

The spatial part of the basis functions with natural parity, $p=(-1)^N$, is constructed using
explicitly correlated Gaussian functions\cite{Bo60,Si60,JeSz79,CeRy93,Ry03}
and the global vector representation\cite{SuUsVa98,VaSuUs98,SuVaBook98} as
\begin{align}
  \phi^{[\lambda]}(\mx{r};\mx{\alpha},\mx{u},K)
  =
  |\mx{v}|^{2K+N} Y_{NM_N}(\hat{\mx{v}})
  \exp\left(%
    -\frac{1}{2}
    \sum_{i=1}^{\npart+1}
    \sum_{j>i}^{\npart+1}
      \alpha_{ij}(\mx{r}_i-\mx{r}_j)^2
  \right) \ ,
  \label{eq:spbasis}
\end{align}
where the $\hat{\mx{v}}=\mx{v}/|\mx{v}|$ unit vector points in the direction of the global vector,
$\mx{v} = \sum\displaystyle_{i=1}^{\npart+1} u_i^{(0)} \mx{r}_i$.
$Y_{NM_N}$ denote the $N$th order $M_N$th degree spherical harmonic function.
The spin function, $\chi^{[\varsigma]}$, is constructed from elementary spin functions so
that the resulting function is an eigenfunction of $\hat{S}_a^2$ and $(\hat{S}_a)_z$
for each type of particles ($a,b,\ldots$) with the quantum numbers $\varsigma=(S_a,M_{S_a},S_b,M_{S_b},\ldots)$.

Then, the resulting $\Phi^{[\lambda,\varsigma]}$ basis function has the quantum numbers of the
non-relativistic quantum theory (it is ``symmetry-adapted'') and contains free parameters,
which can be optimized for an efficient description of the ``internal structure'' of a system.
The free parameters of the spatial function, \ref{eq:spbasis},
are
$K$,
$\mx{\alpha}$: $\alpha_{ij}\ (i=1,\ldots,\npart+1,j=i+1,\ldots,\npart+1)$,
and
$\mx{u}$: $u^{(0)}_i\ (i=1,\ldots,\npart+1)$ with the restriction $\sum_{i=1}^{\npart+1}u^{(0)}_i=0$,
which guarantees translational invariance.
The spin functions used in this work do not contain any free parameters.

We only note here that for the positronium molecule, Ps$_2$, as a special case studied in this work,
the entire basis function was additionally adapted to the
charge-conjugation symmetry of the electrons and the positrons.\cite{Schrader04a,Schrader04b,Schrader07}

The matrix elements of the kinetic and potential energy operators
corresponding to the basis functions with natural parity, \ref{eq:basis}--\ref{eq:spbasis},
were evaluated with the pre-BO program according to Ref.~\citenum{MaRe12}.

Since the basis functions are not orthogonal
we have to solve a generalized eigenvalue problem
\begin{align}
  \mx{H}\mx{c}_i = E_i\mx{S}\mx{c}_i \ ,
  \label{eq:rgeiv}
\end{align}
to find the variationally optimal linear combination of the basis functions
with the linear combination coefficients $\mx{c}_i$ corresponding to the eigenvalue $E_i$.
The generalized eigenproblem is solved by introducing
$\mx{H}' = \mx{S}^{-1/2}\mx{H}\mx{\mx{S}^{-1/2}}$ and $\mx{c}'_i = \mx{S}^{1/2}\mx{c}_i$,
which simplifies the eigenvalue equation, \ref{eq:rgeiv},
to $\mx{H}'\mx{c}'_i=  E_i\mx{c}'_i$.
In our computations the Cholesky decomposition of $\mx{S}$ for the evaluation of $\mx{S}^{-1/2}$
as well as the diagonalization of the real, symmetric transformed Hamiltonian matrix, $\mx{H}'$,
were carried out by using the LAPACK library routines.\cite{lapack}

The computational efficiency and usefulness of the variational approach described depends
on the parameterization of the basis functions
(the choice of the values of $K_I$, $\mx{\alpha}_I$, $\mx{u}_I$ for the basis functions $I=1,\ldots,\nbas$).
For bound-state calculations
we adopted the stochastic variational approach\cite{KuKr77,AlMoSz86,AlMoSz87,AlMoSz88,SuVaBook98}
for the optimization of the basis function parameters.
The (quasi-)optimization of the parameter set is a very delicate problem
and our recipe includes the following details:\cite{MaRe12}
(a) a system-adapted random number generator is constructed using a sampling-importance resampling strategy
    for the generation of trial parameter sets;
(b) the acceptance criterion of the generated trial values is based on
    the linear independence condition and
    the energy minimization requirement (relying on the variational principle);
(c) the selected parameters are fine-tuned using a simple random walk approach or Powell's method\cite{Po04}.

Furthermore, once a parameter set has been selected or optimized for a system with some quantum numbers
it can be used, ``transferred'', to parameterize basis functions for the same system with different quantum numbers
(``parameter transfer approach''). It is important to emphasize that
the basis functions are not transferred, since they have different mathematical form for different
quantum numbers, but only the parameter set is taken over from one calculation to another.

\subsection{Calculation of resonances}
The bound-state pre-BO approach described was extended for
the calculation of resonance states as follows.
First of all, without any change of the computer program,
we looked for the quasi-stabilization of higher-energy eigenvalues
(higher than the lowest-energy threshold) of the real eigenvalue problem.
This application of the stabilization technique\cite{HaTa70,Ho83,UsSu02} 
is a simple, practical test for identifying
possible quasi-bound states and was found to
be useful as a first check of the higher-energy eigenspectrum.
By making full use of the stabilization theory both 
the resonance energies and widths
could be calculated from consecutive diagonalization of the (real) Hamiltonian matrix
corresponding to increasing number of basis functions, 
which cover increasing boxes of the configuration space.\cite{MaRaTa93,MaTaRyMo94,MuYaBu94,RyMoMaTa94} 

Instead of using this approach,
the complex coordinate rotation method (CCRM) \cite{AgCo71,BaCo71,Si72,Rein82,Moi98} was implemented
for the calculation of resonance parameters, energies and widths.
The resonance parameters were determined by identifying stabilization points
in the complex plane with respect to the coordinate rotation angle and
(the size and parameterization of) the basis set.
The localized real part of the eigenvalue, $\mr{Re}(\mathcal{E})$, was taken to be the resonance energy,
and the imaginary part provided the resonance width, $\Gamma=-2\mr{Im}(\mathcal{E})$, which
is inversely proportional to the lifetime, $\tau=\hbar/\Gamma$.\cite{KuKrHo88}

\paragraph{Implementation of the complex coordinate rotation method for the Coulomb Hamiltonian}
  The complex scaling of the coordinates $r \rightarrow r\mr{e}^{\mr{i}\theta}$
  translates to the replacement of the Hamiltonian
  $\hat{H} = \hat{T} + \hat{V}$ with
  \begin{align}
    \hat{\mathcal{H}}(\theta)
    =
    \mr{e}^{-2\mr{i}\theta} \hat{T}
    +
    \mr{e}^{-\mr{i}\theta} \hat{V} \ .
    \label{eq:Htheta}
  \end{align}
  The matrix representation of $\hat{\mathcal{H}}(\theta)$
  is constructed with the matrices of $\hat{T}$ and $\hat{V}$ evaluated by
  the pre-BO program \cite{MaRe12}. Then, the eigenvalue equation for $\hat{\mathcal{H}}(\theta)$
  reads as
  \begin{align}
    \mx{\mathcal{H}}(\theta) \mx{v}_i(\theta) = \mathcal{E}_i(\theta) \mx{S}\mx{v}_i(\theta)  \ ,
    \label{eq:cHop}
  \end{align}
  where $\mx{S}$ is the overlap matrix of the (linearly independent) set of basis functions.
  $\mx{S}$ is eliminated from the equation similarly to
  the case of the real generalized eigenproblem, \ref{eq:rgeiv},
  \begin{align}
    \mx{\mathcal{H}}'(\theta) \mx{v}'_i(\theta) = \mathcal{E}_i(\theta) \mx{v}'_i(\theta) \ ,
    \label{eq:cHpmx}
  \end{align}
  with
  \begin{align}
    \mx{\mathcal{H}}'(\theta)
    &=
    \mr{e}^{-2\mr{i}\theta} \mx{T}'
    +
    \mr{e}^{-\mr{i}\theta} \mx{V}' \nonumber \\
    &=
     \cos(2\theta) \mx{T}' + \cos(\theta)\mx{V}'
    -\mr{i}(\sin(2\theta)\mx{T}' + \sin(\theta)\mx{V}')
    \label{eq:cHpmx2}
  \end{align}
  and
  \begin{align}
    \mx{T}' = \mx{S}^{-1/2}\mx{T}\mx{S}^{-1/2}
    \quad\text{and}\quad
    \mx{V}' = \mx{S}^{-1/2}\mx{V}\mx{S}^{-1/2} \ .
  \end{align}
  The complex symmetric eigenvalue problem, \ref{eq:cHpmx}, is solved
  using the LAPACK library routines.\cite{lapack}

%
%
\clearpage
\section{Numerical results\label{ch:numres}}
The first numerical applications of our implementation were carried out for
the notoriously non-adiabatic positronium anion, Ps$^-=\{\mr{e}^-,\mr{e}^-,\mr{e}^+\}$, and
for the positronium molecule, Ps$_2=\{\mr{e}^-,\mr{e}^-,\mr{e}^+,\mr{e}^+\}$.
The reason for the choice of these systems was of technical nature:
we observed in bound-state calculations \cite{MaRe12} that it was straightforward
to find an appropriate parameterization of the basis set for the positronium complexes.
Furthermore, comparison of the results with earlier calculations\cite{Papp02,LiSh05,UsSu02,SuUs04} allowed us
check the developed computational methods and gain experience
in the localization of the real and imaginary parts of the complex eigenvalues
using the complex coordinate rotation method.

While for the bound-state calculations the basis function parameters were optimized for the
lowest-energy level(s) using the variational principle, this handy optimization
criterion was not available in the CCRM calculations. Thus, we used optimized bound-state basis sets
and enlarged them with linearly independent basis functions for the estimation of the resonance parameters.

Next, we investigated the calculation of some of the excited states of the H$_2$ molecule.
The construction of a reasonably good parameterization for the basis set
has turned out to be a challenge.
Nevertheless, we describe the essence of our observations and
give calculated resonance energy values and approximate widths
for the lowest-lying excited states beyond the $\b3Sup$ repulsive electronic state embedded
in the H(1)+H(1) continuum.

\subsection{Ps$^-$\label{sec:Psm}}
In \ref{tab:resPsm} bound- and resonance-state parameters ($N=0$, $p=+1$) are collected,
which were obtained in this work. 
The basis sets were generated using the energy minimization and the linear independence conditions
using a random number generator set up following the strategy described in Ref.~\citenum{MaRe12}.
The parameters for the largest basis sets used during the calculations are given in the \som.
The generated basis sets were apparently large and flexible enough to obtain
resonance states not only beyond the first, but also beyond the second, and the third dissociation channels
which correspond to $\mr{Ps}(1)+\mr{e}^-$, $\mr{Ps}(2)+\mr{e}^-$, and $\mr{Ps}(3)+\mr{e}^-$, respectively.
As to the accuracy, the (real) variational principle, directly applicable for bound states, 
is not useful for the assessment of the resonance parameters.
Instead, we used benchmarks available in the literature 
resulting from extensive three-body calculations 
using Pekeris-type wave functions with one and two length scales\cite{LiSh05}
and from the solution of the Fadeev--Merkuriev integral equations for three-body systems.\cite{Papp02}

First of all, the present results and the literature data are in satisfactory agreement.
Our results could be certainly improved by running more extensive calculations with larger basis sets.
Instead of going in this direction, a careful comparison is carried out with the reference data
in order to learn about the accuracy and convergence behavior of our approach.
The results are often in excellent agreement with the benchmarks
but in some cases the lifetimes are orders of magnitude off. 
It can be observed that the calculated lifetimes are
worse when the real part of the complex energy was determined less accurately
(and given to less significant digits in \ref{tab:resPsm}). 
The inaccuracy appears in both the real and the imaginary parts 
and is about of the same order of magnitude compared to the absolute value of the complex energy.
Thus, if the widths are expected to be very small and the real part
can be determined only to a few digits, the width should be considered
only as a rough estimate to its exact value.
This observation can be used later in this work also 
for the assessment of the calculations carried out for
the four-particle Ps$_2$ and H$_2$.

\begin{table}
  \caption{%
    Identified bound- and resonance-state energies and resonance widths, in \Eh,
    of Ps$^-=\lbrace \mr{e}^-,\mr{e}^-,\mr{e}^+\rbrace$.$^\mr{a}$
    \label{tab:resPsm}
    ~\\[-1.cm]
  }
  \begin{center}
    \begin{tabular}{@{}c l@{\ \ \ }r l@{\ \ \ }r r@{}}
      \cline{1-6} \\[-0.4cm]
      \cline{1-6} \\[-0.3cm]
      \multicolumn{1}{c}{$(N,p,S_-)$ $^\mr{b}$} &
      \multicolumn{1}{c}{$\mr{Re}(\mathcal{E})$ $^\mr{c}$} &
      \multicolumn{1}{c}{$\Gamma/2$ $^\mr{c}$} &
      \multicolumn{1}{c}{$\mr{Re}(\mathcal{E}_\mr{Ref})$ $^\mr{d}$} &
      \multicolumn{1}{c}{$\Gamma_\mr{Ref}/2$ $^\mr{d}$} &
      Ref. \\
      \cline{1-6} \\[-0.3cm]
      $(0,+1,0)$ & $-0.262\ 005\ 070$\ $^\mr{e}$ & $0$\ $^\mr{e}$
                 & $-0.262\ 005\ 070$            & $0$            & \cite{Ko00} \\[0.15cm]
      \cdashline{1-6} \\[-0.25cm]
      $(0,+1,0)$ & $-0.076\ 030\ 455$ & $2.152 \cdot 10^{-5}$
                 & $-0.076\ 030\ 442$ & $2.151\ 7 \cdot 10^{-5}$ & \cite{LiSh05} \\
      $(0,+1,0)$ & $-0.063\ 649\ 173$ & $4.369 \cdot 10^{-6}$
                 & $-0.063\ 649\ 175$ & $4.339\ 3 \cdot 10^{-6}$ & \cite{LiSh05} \\
      $(0,+1,0)$ & $-0.062\ 609$      & $2.5 \cdot 10^{-5}$
                 & $-0.062\ 550$      & $5.0 \cdot 10^{-7}$      & \cite{Papp02} \\[0.15cm]
      \cdashline{1-6} \\[-0.25cm]
      $(0,+1,0)$ & $-0.035\ 341\ 85$ & $3.730 \cdot 10^{-5}$
                 & $-0.035\ 341\ 885$ & $3.732\ 9 \cdot 10^{-5}$ & \cite{LiSh05} \\
      %
      $(0,+1,0)$ & $-0.029\ 845\ 70$ & $2.781 \cdot 10^{-5}$
                 & $-0.029\ 846\ 146$ & $2.635\ 6 \cdot 10^{-5}$ & \cite{LiSh05} \\
      $(0,+1,0)$ & $-0.028\ 271$      & $1.8 \cdot 10^{-5}$
                 & $-0.028\ 200$      & $7.5 \cdot 10^{-6}$      & \cite{Papp02} \\[0.15cm]
      \cdashline{1-6} \\[-0.25cm]
      $(0,+1,0)$ & $-0.020\ 199$ & $8.800 \cdot 10^{-5}$
                 & $-0.020\ 213\ 921$ & $6.502\ 6 \cdot 10^{-5}$ & \cite{LiSh05} \\
      \cline{1-6} \\[-0.3cm]
      \cdashline{1-6} \\[-0.25cm]
      $(0,+1,1)$ & $-0.063\ 537\ 352$ & $2.132 \cdot 10^{-9}$
                 & $-0.063\ 537\ 354$ & $1.570\ 0 \cdot 10^{-9}$ & \cite{LiSh05} \\
      $(0,+1,1)$ & $-0.062\ 591$      & $2.6 \cdot 10^{-7}$
                 & $-0.062\ 550$      & $2.5 \cdot 10^{-10}$     & \cite{Papp02} \\[0.15cm]
      \cdashline{1-6} \\[-0.25cm]
      $(0,+1,1)$ & $-0.029\ 369\ 87$ & $1.300 \cdot 10^{-7}$
                 & $-0.029\ 370\ 687$ & $9.395\ 0 \cdot 10^{-8}$ & \cite{LiSh05} \\

      $(0,+1,1)$ & $-0.028\ 21$       & $1.9 \cdot 10^{-5}$
                 & $-0.028\ 05$       & $5.0 \cdot 10^{-8}$      & \cite{Papp02} \\[0.15cm]
      \cdashline{1-6} \\[-0.25cm]
      $(0,+1,1)$ & $-0.017\ 071$ & $6.710 \cdot 10^{-6}$
                 & $-0.017\ 101\ 172$ & $3.560\ 9 \cdot 10^{-7}$ & \cite{Papp02} \\
      \cline{1-6} \\[-0.4cm]
      \cline{1-6} \\[-0.3cm]
    \end{tabular}
  \end{center}
  \begin{flushleft}
    ~\\[-0.5cm]
    $^\mr{a}$ %
      The dissociation threshold energies, in \Eh, accessible for both the $S_-=0$ and 1 states are
      $E(\mr{Ps}(1))=-1/4=-0.25$,
      $E(\mr{Ps}(2))=-1/16=-0.062\ 5$,
      and
      $E(\mr{Ps}(3))=-1/36=-0.027\ \dot{7}$.\\
    $^\mr{b}$ %
      $N,p,$ and $S_-$: total spatial angular momentum quantum number, parity, and
      total spin quantum number of the electrons, respectively. \\
    $^\mr{c}$ %
      $\mr{Re}(\mathcal{E})$ and $\Gamma$:
      resonance energy and width with $\Gamma/2=-\mr{Im}(\mathcal{E})$ calculated in this work. \\
    $^\mr{d}$ %
      $\mr{Re}(\mathcal{E}_\mr{Ref})$ and $\Gamma_\mr{Ref}$:
      resonance energy and width with $\Gamma/2=-\mr{Im}(\mathcal{E}_\mr{Ref})$ taken from 
      Refs.~\citenum{Papp02} and \citenum{LiSh05}. \\
    $^\mr{e}$ %
      Bound state.
  \end{flushleft}
\end{table}

\clearpage
\subsection{Ps$_2$\label{sec:Ps2}}
Our next test case was the four-particle positronium molecule,
Ps$_2=\lbrace \mr{e}^-,\mr{e}^-,\mr{e}^+,\mr{e}^+ \rbrace$.
Resonances of the positronium molecule have recently attracted attention \cite{DiDr10,SuUs04,UsSu02}.
Ps$_2$ has few bound states, and thus a detailed spectroscopic investigation
of its structure and dynamics is possible only through
the detection and analysis of its quasi-bound states.

In our list of numerical examples,
the positronium molecule is unique because, in addition to the spatial symmetries,
its Hamiltonian is invariant under the conjugation of the electric charges.
In order to account for this additional property,
the basis functions, \ref{eq:basis}--\ref{eq:spbasis}, were adapted
also to the charge conjugation symmetry.\cite{SuUs00,SuUs04,Schrader04a,Schrader04b,Schrader07}
As a result, the total symmetry-adapted basis functions and also the calculated wave functions are not necessarily
eigenfunctions for the total spin angular momentum of the electrons or that of the positrons,
$\hat{S}^2_-$ or $\hat{S}^2_+$. Nevertheless, the total spatial angular momentum quantum number, $N$,
the parity, $p$, as well as the charge-conjugation parity, $c=+1$ or $-1$, are always good quantum numbers.

The parameterization strategy of the basis set was similar to that used for Ps$^-$:
we employed (a) the energy minimization condition for the lowest-energy eigenvalue;
and (b) the linear independence condition for the generation of new basis function parameters.
The parameter sets used in the largest calculations are given in the \som.

\begin{figure}
  \includegraphics[scale=1.]{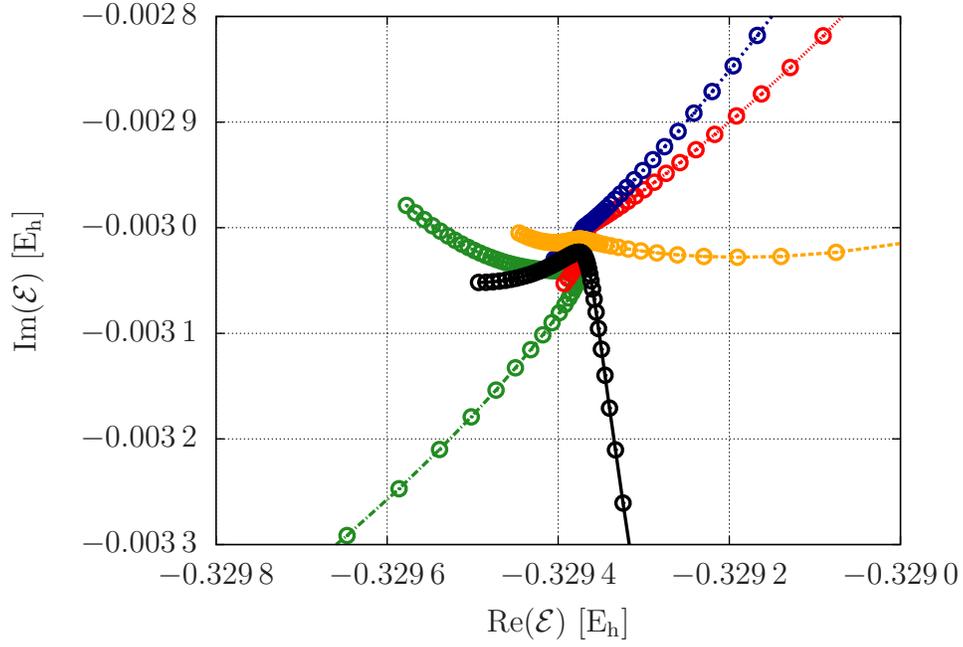}
  \caption{%
    Localization of the parameters for the lowest-energy resonance state of
    $\mr{Ps}_2$ with $N=0$, $p=+1$, $c=+1$, and $S_\mr{-}=0$, $S_\mr{+}=0$.
    The stabilization of the trajectories with respect to the rotation angle (circles) and
    the basis functions (colors) is shown. The stabilization
    point is located at $(\mr{Re}(\mathcal{E}),\mr{Im}(\mathcal{E})) = (-0.329\ 38,-0.003\ 03)$~\Eh.
    \label{fig:Ps2res1}
  }
\end{figure}

The bound and resonance states calculated with $N=0$ total spatial angular momentum quantum number
and $p=+1$ parity are collected in \ref{tab:resonPs2}. Considering all possible charge conjugation and spin functions,
we obtained only three bound states in agreement with the literature.\cite{BuAd06,UsSu02,SuUs04} Two of the three calculated bound
states substantially improve on the best available results.\cite{SuUs04}
It is interesting to note that the bound state
with $E=-0.314\ 677\ 072$~\Eh\ ($c=-1$ and $(S_-,S_+)=(1,1)$) is bound
only due to the charge-conjugation symmetry of the electrons and the positrons.
The localization of the energy and width for the lowest-energy resonance state of Ps$_2$
is shown in \ref{fig:Ps2res1}.
The calculated resonance positions are in good agreement with the literature.
In some cases our results might even improve on the best available data,\cite{UsSu02,SuUs04}
although there is no such a direct criterion for the assessment of the accuracy of 
the resonance parameters as the variational principle for bound states.

\begin{table}
  \caption{%
    Identified bound- and resonance-state energies and resonance widths, in \Eh, of
    Ps$_2=\lbrace \mr{e}^-,\mr{e}^-,\mr{e}^+,\mr{e}^+\rbrace$.$^\mr{a}$
    \label{tab:resonPs2}
    ~\\[-1.cm]
  }
  \begin{center}
    \begin{tabular}{@{}c@{ }c l@{\ \ }r l@{\ \ }r r@{}}
      \cline{1-7} \\[-0.4cm]
      \cline{1-7} \\[-0.3cm]
      \multicolumn{1}{c}{$(N,p,c)$ $^\mr{b}$} &
      \multicolumn{1}{c}{$(S_-,S_+)$ $^\mr{c}$} &
      \multicolumn{1}{c}{$\mr{Re}(\mathcal{E})$ $^\mr{d}$} &
      \multicolumn{1}{c}{$\Gamma/2$ $^\mr{d}$} &
      \multicolumn{1}{c}{$\mr{Re}(\mathcal{E}_\mr{Ref})$ $^\mr{e}$} &
      \multicolumn{1}{c}{$\Gamma_\mr{Ref}/2$ $^\mr{e}$} &
      \multicolumn{1}{c}{Ref.} \\
      \cline{1-7} \\[-0.3cm]
      %
      $(0,+1,+1)$ & $(0,0)$ &
      $-0.516\ 003\ 789\ 741$\ $^\mr{f}$ & $0$\ $^\mr{f}$        &
      $-0.516\ 003\ 790\ 416$            & $0$                   & \cite{BuAd06} \\ 
      $(0,+1,+1)$ & $(0,0)$ &
      $-0.329\ 38$                       & $3.03 \cdot 10^{-3}$  &
      $-0.329\ 4$                        & $3.1 \cdot 10^{-3}$   & \cite{SuUs04} \\ 
      $(0,+1,+1)$ & $(0,0)$ &
      $-0.291\ 7$                        & $2.5 \cdot 10^{-3}$   &
      $-0.292\ 4$                        & $1.95 \cdot 10^{-3}$  & \cite{SuUs04} \\ 
      \cline{1-7} \\[-0.3cm]
      \colgray
      $(0,+1,-1)$ & $(0,0)$ &
      $-0.314\ 677\ 072$\ $^\mr{f}$      & $0$\ $^\mr{f}$          &
      $-0.314\ 673\ 3$                   & $0$                     & \cite{SuUs04} \\ 
      $(0,+1,-1)$ & $(0,0)$ &
      $-0.289\ 789\ 3$                   & $7.7 \cdot 10^{-5}$     &
      $-0.289\ 76$                       & $7 \cdot 10^{-5}$       & \cite{SuUs04} \\ 
      $(0,+1,-1)$ & $(0,0)$ &
      $-0.279\ 25$                       & $2.3 \cdot 10^{-4}$     &
      $-0.279\ 13$                       & $1 \cdot 10^{-4}$       & \cite{SuUs04} \\ 
      %
      %
      \colblack\\[-0.5cm]
      \cline{1-7} \\[-0.3cm]
      \colgray
      $(0,+1,+1)$ & $(1,1)$ &
      $-0.277\ 2$                        & $5.4 \cdot 10^{-4}$   &
      $-0.276\ 55$                       & $1.55 \cdot 10^{-4}$  & \cite{SuUs04} \\ 
      \colblack\\[-0.5cm]
      \cline{1-7} \\[-0.3cm]
      $(0,+1,-1)$ & $(1,1)$ &
      $-0.309\ 0$                        & $5.7 \cdot 10^{-3}$   &
      $-0.308\ 14$                       & $1.2 \cdot 10^{-4}$   & \cite{SuUs04} \\ 
      $(0,+1,-1)$ & $(1,1)$ &
      $-0.273\ 3$                        & $2.3 \cdot 10^{-3}$   &
      $-0.273\ 6$                        & $8.5 \cdot 10^{-4}$   & \cite{SuUs04} \\ 
      %
      \cline{1-7} \\[-0.3cm]
      \colgray
      $(0,+1,\pm 1)$ & $(1,0)/(0,1)$ &
      $-0.330\ 287\ 505$\ $^\mr{f}$      & $0$\ $^\mr{f}$          &
      $-0.330\ 276\ 81$                  & $0$                     & \cite{SuUs04} \\ 
      $(0,+1,\pm 1)$ & $(1,0)/(0,1)$ &
      $-0.294\ 3$                        & $3.1  \cdot 10^{-3}$    &
      $-0.293\ 9$                        & $2.15 \cdot 10^{-3}$    & \cite{SuUs04} \\ 
      $(0,+1,\pm 1)$ & $(1,0)/(0,1)$ &
      $-0.282$                           & $2 \cdot 10^{-3}$       &
      $-0.282\ 2$                        & $8.5 \cdot 10^{-4}$     & \cite{SuUs04} \\ 
      \colblack\\[-0.5cm]
      \cline{1-7} \colblack\\[-0.4cm]
      \cline{1-7} \colblack\\[-0.3cm]
    \end{tabular}
  \end{center}
  \begin{flushleft}
    ~\\[-0.5cm]
    \colblack
    $^\mr{a}$~%
      For the five symmetry blocks with different $(N,p,c)$ quantum numbers and
      $(S_-,S_+)$ labels the lowest accessible thresholds are
      Ps(1S)+Ps(1S),
      {\colgray Ps(1S)+Ps(2P),}
      {\colgray Ps(1S)+Ps(2P),}
      Ps(1S)+Ps(1S),
      {\colgray Ps(1S)+Ps(2S,2P),} respectively.\cite{SuUs00}
      The corresponding energies, in \Eh, are
      $E(\mr{Ps(1)+Ps(1)})=-1/2=-0.5$ and
      {\colgray $E(\mr{Ps(1)+Ps(2)})=-5/16=-0.312\ 5$.}
      (The black and gray coloring is used to help the orientation.) \\
    $^\mr{b}$~%
      $N,p,$ and $c$: total spatial angular momentum quantum number,
      parity, and charge conjugation quantum number, respectively. \\
    $^\mr{c}$~%
      $S_-$ and $S_+$: total spin quantum numbers for the electrons and the positrons, respectively.
      In the last symmetry block, $(S_-,S_+)=(0,1)$ and $(S_-,S_+)=(1,0)$, are not good quantum numbers
      because these spin states are coupled due to the charge-conjugation symmetry of the Hamiltonian. \\
    $^\mr{d}$ %
      $\mr{Re}(\mathcal{E})$ and $\Gamma$:
      resonance energy and width with $\Gamma/2=-\mr{Im}(\mathcal{E})$ calculated in this work. \\
    $^\mr{e}$ %
      $\mr{Re}(\mathcal{E}_\mr{Ref})$ and $\Gamma_\mr{Ref}$:
      resonance energy and width with $\Gamma/2=-\mr{Im}(\mathcal{E}_\mr{Ref})$ taken from Ref.~\citenum{SuUs04}. \\
    $^\mr{f}$~%
      Bound states.
  \end{flushleft}
\end{table}

\clearpage
\subsection{Toward the calculation of rovibronic resonances of H$_2$}
Next, our goal was to explore how the lowest-lying resonance states of H$_2$
can be calculated in a pre-Born--Oppenheimer quantum mechanical approach.
It could be anticipated that one of the major challenges in this undertaking
would be the parameterization of the basis functions,
which was already in the bound-state calculations more demanding for H$_2$ than for Ps$_2$.\cite{MaRe12}
In the bound-state calculations the optimized parameters were fine-tuned in repeated cycles.
The entire parameter selection and optimization procedure relied on the variational principle
and the minimization of the energy.

\begin{figure}
  \includegraphics[width=0.9\linewidth]{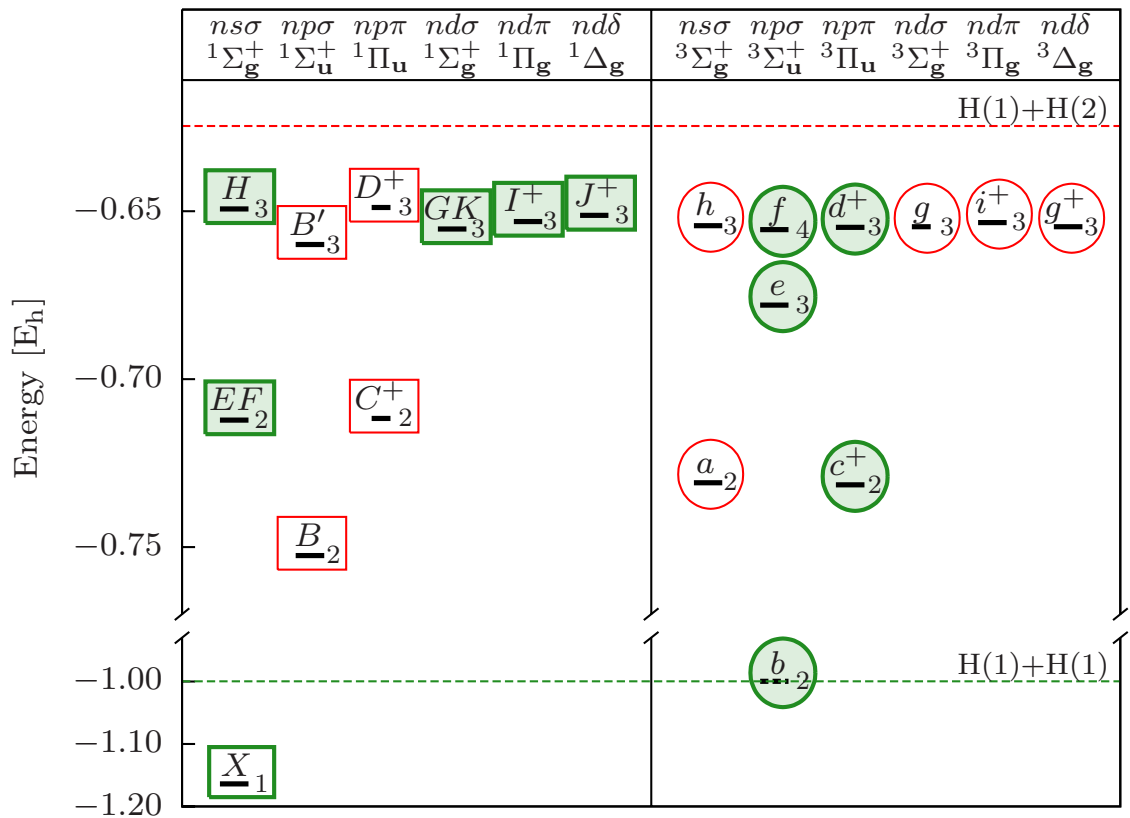}
  \caption{%
    Orientation chart for the electronic states of H$_2$ below the H(1)+H(2) dissociation threshold
    (see for example Herzberg\cite{Herzberg1} or Brown and Carrington\cite{BrCa03}).
    The same color (red or green) and shape (rectangle or ellipse) coding indicate those states, which
    can be obtained in the same pre-Born--Oppenheimer calculation. Empty objects indicate bound states, while
    filled objects refer to the fact that the corresponding rovibronic states (if there are any)
    are resonances embedded in the H(1)+H(1) continuum.
    \label{fig:orientH2}
  }
\end{figure}

According to the spatial and permutational symmetry properties of the H$_2$ molecule,
there are four different blocks with natural parity
\begin{itemize}
  \item[B1: ]
    ``$\X1Sgp$ block'': $N\geq 0,\ p=(-1)^N,\ \Sp=(1-p)/2,\ \Se=0$;
  \item[B2: ]
    ``$\B1Sup$ block'': $N\geq 0,\ p=(-1)^N,\ \Sp=(1+p)/2,\ \Se=0$;
  \item[B3: ]
    ``$\a3Sgp$ block'': $N\geq 0,\ p=(-1)^N,\ \Sp=(1-p)/2,\ \Se=1$;
  \item[B4: ]
    ``$\b3Sup$ block'': $N\geq 0,\ p=(-1)^N,\ \Sp=(1+p)/2,\ \Se=1$,
\end{itemize}
which can be calculated in independent runs with our computer program using basis functions with
the appropriate quantum numbers, \ref{eq:basis}--\ref{eq:spbasis}.
The lowest-energy levels of the first three blocks correspond to bound states,
while the last block starts with the H(1)+H(1) continuum.
In the BO picture the $\b3Sup$ electronic state is repulsive \cite{KoRy90}
and does not support any bound rotational-vibrational levels
(see \ref{fig:orientH2} and \ref{tab:boundH2}).
Then, one of our goals was the identification of
the lowest-energy quasi-bound states in the $\b3Sup$ block.

During the calculation of the lowest-energy resonances of Ps$_2$,
the basis function parameters were generated randomly
using a system-adapted random number generator.\cite{MaRe12}
Unfortunately, this simple strategy for the H$_2$ resonances was not useful.

The energy minimization criterion for the lowest (few) eigenstates was not useful neither,
since it resulted only in the accumulation of functions near the H(1)+H(1) limit,
the lowest energy levels in the $\b3Sup$ block,
and the higher-lying quasi-bound states were not at all described by
the basis sets generated in this way.

Then, our alternative working strategy was the usage of the parameter transfer approach
(described in Section ``Theory and computational strategy'' and in Ref.~\citenum{MaRe12}).
In this approach a parameter set optimized for a bound state with some quantum numbers, ``state $\mathcal{A}$'',
is used to parameterize the basis functions corresponding to another
set of quantum numbers and used to calculate ``state $\mathcal{B}$''.
It is important to emphasize that the mathematical form of the basis functions is defined by the selected
values of the quantum numbers, \ref{eq:basis}--\ref{eq:spbasis}, and thus not the basis functions but only
the parameters are transferred from one calculation to another.
Our qualitative understanding tells us that this parameter-transfer strategy is computationally useful
if the internal structures of ``state $\mathcal{A}$'' and ``state $\mathcal{B}$'' are more or less similar.
By inspecting the orientation chart of H$_2$, \ref{fig:orientH2}, our idea was that the combination of the
(natural-parity) bound-state optimized parameter sets could provide a parameterization good enough
for the identification of the lowest-lying resonance states embedded in the $\b3Sup$ continuum.

For this purpose, we used the parameters of $2\ 250$ basis functions optimized for the lowest-lying bound states with
$N=0$ and $1$ angular momentum quantum numbers corresponding to the $\X1Sgp$, $\B1Sup$, and $\a3Sgp$ blocks
using the sampling-importance resampling strategy of Ref.~\citenum{MaRe12} and Powell's method\cite{Po04}
for the fine-tuning of each basis function.
As a result of these calculations, we obtained
a parameter set large enough for $6\times 2\ 250=13\ 500$ basis functions.
In addition, $1\ 000-1\ 000$ basis functions were generated and less tightly optimized for
the lowest-energy levels of the $\b3Sup$ block with $N=0$ and 1.
Using this large parameter set, $\mathcal{P}_L$,
$15\ 500$ basis functions were constructed for all possible quantum numbers of
the four blocks, B1--B4, with $N=0,1,$ and 2.
In each case the resulting basis set was found to be linearly independent.
The complete parameter set is given in the \som.
The proton-electron ratio was $m_\mr{p}/m_\mr{e}=1\,836.152\,672\,47$ throughout the calculations.\cite{codata06}

\vspace{0.1cm}
\paragraph{Bound-state energy levels}
The lowest-lying energy values obtained with $\mathcal{P}_L$ for the different quantum numbers
are collected in \ref{tab:boundH2},
and are in good agreement with the best available non-relativistic results in the literature.
The energy values of the $\X1Sgp$ electronic and vibrational ground states with
$N=0,1,$ and $2$ are larger by only less than $2$~n\Eh\ than the theoretical results of Ref.~\citenum{PaKo09}.

For all three calculated $\B1Sup$ $N=0,1,2$ levels our energy values
are lower by more than $1\ \mu\mr{E}_\mr{h}$ compared to the results
of Ref.~\citenum{WoOrSt06} obtained in close-coupling calculations
using adiabatic potential energy curves and non-adiabatic couplings for six
electronic states.
We also note that for the $N=0$ lowest-lying vibrational level of
$\B1Sup$ there is a ``variational-perturbational'' estimate given
in Table~3 of Ref.~\citenum{WoOrSt06}, 
which was anticipated to be more accurate, and thus it was the recommended value for this level 
in the article, though not a strict upper bound to 
the exact value.
It was obtained not in the six-state close-coupling calculations,
but as a result of two-state close-coupling calculations with the
potentials and couplings of Ref.~\citenum{WoOrSt06}, incremented with a
non-adiabatic correction term.\cite{WoDr92} This
term value translates to the energy value $-0.753\ 026\ 440$\ \Eh\
based on the explanation given below Eq. (13) of
Ref.~\citenum{WoOrSt06}. 
An earlier non-adiabatic estimate\cite{Wo95} (not upper bound) for this energy
level was $-0.753\ 027\ 31$\ \Eh\ calculated using the adiabatic
energy and a correction to the BO potential\cite{DrWo95} 
incremented by a non-adiabatic correction.\cite{WoDr92}
For comparison, the pre-Born--Oppenheimer
energy calculated in the present work (in a fully variational
procedure) is $-0.753\ 027\ 186$\ \Eh\ (\ref{tab:boundH2}).

In the case of the $\a3Sgp$ $N=1,2$ energy levels the presented energy values obtained 
in this work are lower than the lowest energies values published\cite{Wo07}.

Based on this overview, we can conclude that the parameter set, $\mathcal{P}_L$, performs
well for the lowest-lying bound-state energy levels and 
also contains basis functions optimized for an approximate
description of the H(1)+H(1) continuum.
Then, we can hope that the application of this parameter set
in the CCRM calculations for the description of the related
or just energetically nearby-lying quasi-bound states will be useful.

\vspace{0.1cm}
\paragraph{Electronically excited bound and resonance rovibronic states}
In the orientation chart of H$_2$, \ref{fig:orientH2}, the electronic states are collected
below the H(1)+H(2) dissociation threshold known from the literature\cite{Herzberg1,BrCa03}
(only natural-parity states are considered).
Although in our calculations there are no potential energy curves corresponding to electronic states,
these conventional electronic-state labels help the orientation and the reference to
the calculated energy levels.
In the figure those states which are coupled by symmetry and calculated in the same
block are highlighted similarly (green or red color and oval or rectangular marking) corresponding to
the B1--B4 blocks introduced earlier this section. This coupling is included in the calculations automatically
by specifying the total spatial (orbital plus rotational) angular momentum,
parity, and spin quantum numbers.
The empty ellipses and rectangles indicate bound states, while the shaded signs are for
resonance states embedded in their corresponding lowest-lying continuum (here: H(1)+H(1)).

We carried out calculations in all four blocks, B1--B4, with $N=0,1,$ and 2 total spatial angular momentum quantum numbers and
most of the states indicated in \ref{fig:orientH2} could have been identified using the largest parameter set,
$\mathcal{P}_L$. Unfortunately, the accuracy of the calculated energies often did not meet the level of
spectroscopic accuracy,\cite{spectracc} and thus we collect here only the essence of the calculations.

\begin{figure}
  \begin{center}
    \includegraphics[scale=1.]{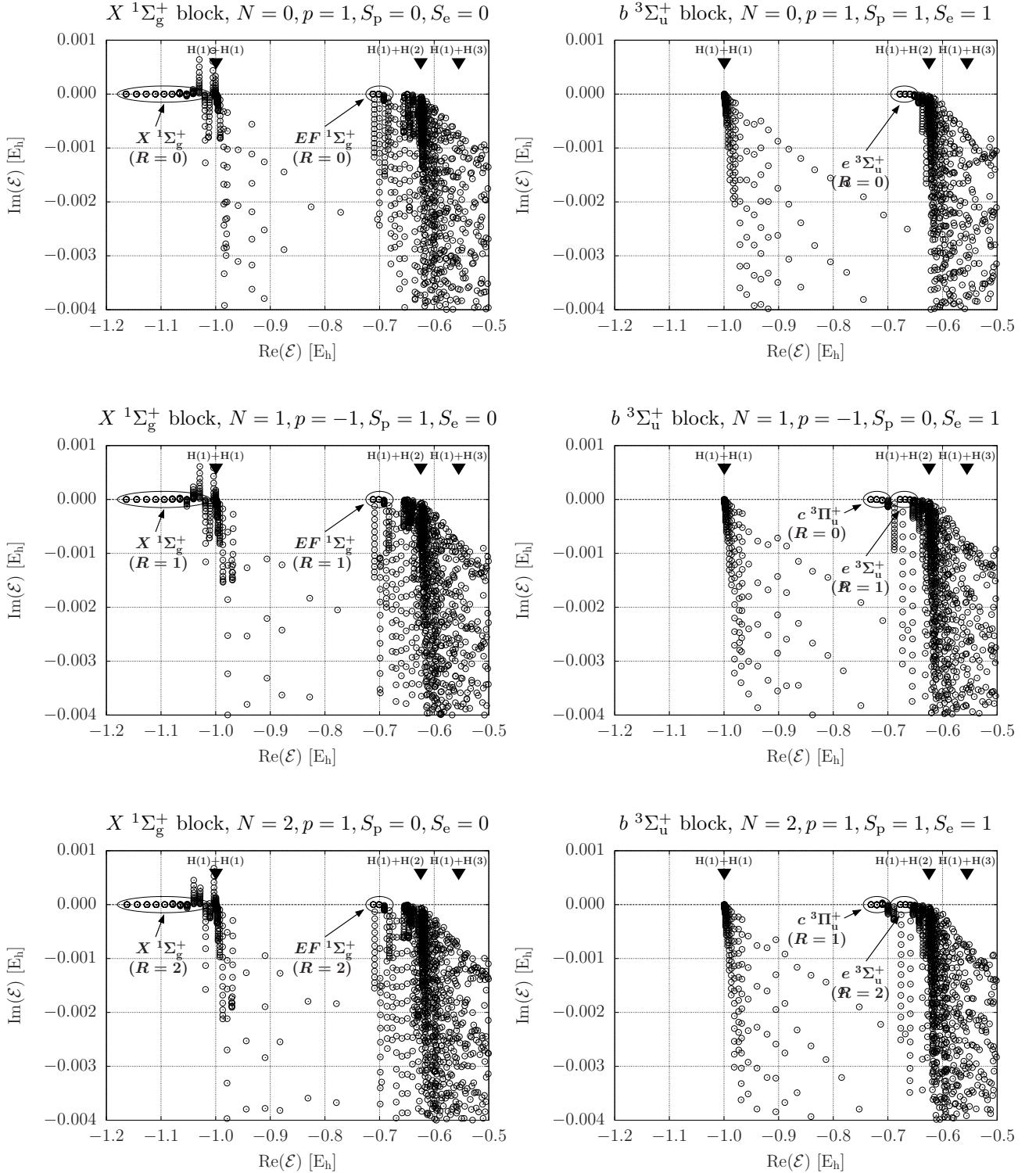} \\
  \end{center}
  \caption{%
    Part of the spectrum of the complex scaled Hamiltonian,
    $\mathcal{H}(\theta)$ with $\theta\in[0.005,0.065]$
    corresponding to the largest basis set used in this work
    for the $X\ ^1\Sigma_\mr{g}^+$ block $[p=(-1)^N,S_\mr{p}=(1-p)/2,S_\mr{e}=0]$
    and
    for the $b\ ^3\Sigma_\mr{u}^+$ block $[p=(-1)^N,S_\mr{p}=(1+p)/2,S_\mr{e}=1]$
    with $N=0,1,$ and $2$ total spatial angular momentum quantum numbers.
    The black triangles indicate the threshold energy of the dissociation continua corresponding to
    H(1)+H(1), H(1)+H(2), and H(1)+H(3).
    \label{fig:resonH2}
  }
\end{figure}

First of all, the most important qualitative results can be explained by inspecting
\ref{fig:resonH2} prepared for the ``$\X1Sgp$ block'' and for the ``$\b3Sup$ block'', B1 and B4.
\ref{fig:resonH2} shows a part of the eigenspectrum of the complex scaled Hamiltonian,
$\mathcal{H}(\theta)$ of \ref{eq:Htheta}, corresponding to small $\theta$ values, $\theta\in[0.005,0.065]$,
and the $[-1.2,-0.5]$~\Eh\ interval of the real part of the eigenvalues.

In both cases the onset of the H(1)+H(1) continuum can be observed on the real axes at $-0.999\ 455\ 679\ \mr{E}_\mr{h}$.
In the $\X1Sgp$ block the bound rovibrational energy levels assignable to the $\X1Sgp$ electronic state line up on
the real axis with $\mr{Im}(\mathcal{E})=0$ (deviations from this value are due to the incomplete convergence only
and the estimated stabilization points are on the real axis).
In the $\b3Sup$ block however we find no states below the H(1)+H(1) continuum, in agreement with the BO calculations \cite{KoRy90}.
The next, H(1)+H(2) threshold corresponds to the energy value $-0.624\ 659\ 800\ \mr{E}_\mr{h}$.
Beyond the H(1)+H(1) but below the H(1)+H(2) thresholds stabilization points (with respect to the $\theta$
rotation angle and the basis set) were observed with small negative imaginary values, which were assigned
(based on the real parts of the eigenvalues) to rotational-vibrational levels
corresponding to the electronically excited
states in their symmetry blocks (see \ref{fig:resonH2} and also \ref{fig:orientH2}).
The lack of any stabilization points beyond the H(1)+H(2) threshold
can be explained with the limited size and flexibility of the basis set.

It can be observed in \ref{fig:resonH2} in the $\b3Sup$ block
that a group of stabilization points appear for $N=1$ and $2$, which are not present for $N=0$.
These points for $N=1$ and $2$ were assigned to the rotational-vibrational states with
$R=0$ and 1 rotational angular momentum quantum numbers of the $c\ ^3\Pi_\mr{u}^+$
electronic state, respectively.
This result demonstrates that the coupling of the rotational and orbital angular
momenta are automatically included in the calculations by specifying only
the total spatial angular momentum quantum number, $N$.

Finally, we note that the H(1)+H(1) continuum does not couple neither to
the ``$\B1Sup$ block'' nor to the ``$\a3Sgp$ block'',
and in these cases the lowest-lying continuum corresponds to
the H(1)+H(2) dissociation channel (\ref{fig:orientH2}).

\vspace{0.1cm}
\paragraph{Numerical results for the resonance energies and widths}
In \ref{tab:resonH2} numerical results are given obtained for the ``$\b3Sup$ block'', B4,
for the lowest-lying rotational-vibrational levels with $N=0,1$, and 2 corresponding to the
$e\ ^3\Sigma_\mr{u}^+$ ($N=0,1,2$) and to the $c\ ^3\Pi_\mr{u}^+$ ($N=1,2$) electronic-state labels.
These rovibronic levels are embedded in the H(1)+H(1) continuum,
and thus they are considered as rovibronic resonances.

The real energy values are in satisfactory agreement with the experimental results.\cite{Di58} 
We consider however the given imaginary parts as estimates to their accurate values,
and all we can conclude at this point is 
that the obtained imaginary parts are of the order of $10^{-7}$~\Eh, which
corresponds to a predissociative lifetime, $\tau=\hbar/\Gamma$, of the order of $0.2$~ns.
As it was explained earlier, 
it is difficult to assess the accuracy of the calculated
resonance energies and widths, since there is no such a simple criterion as 
the (real) variational principle for bound states. 
According to our observations in the test calculations 
for Ps$^-$ (\ref{tab:resPsm}), the relative errors in the real and imaginary parts with respect to 
the absolute value of the complex energy are similar. 
The accuracy of the pre-BO real energy values can be estimated here by their comparison 
with available experimental results. This observation also indicates that
the resonance widths given in \ref{tab:resonH2} should be considered as rough estimates.

Theoretical energy values are available in the literature calculated by Ko{\l}os
and Rychlewski\cite{KoRy77,KoRy90} and are also cited in \ref{tab:resonH2}.
In Ref.~\citenum{KoRy90} adiabatic rotational-vibrational energy levels were determined 
for the $e\ ^3\Sigma_\mr{u}^+$ electronic state by
calculating an accurate adiabatic potential energy curve and solving 
the corresponding rotational-vibrational Schr\"odinger equation.
The theoretical reference value for the $c\ ^3\Pi_\mr{u}^+, R=0, v=0$ level was taken from
Ref.~\citenum{KoRy77}, which was obtained by the calculation of 
an accurate BO potential energy curve and 
solving the corresponding vibrational Schr\"odinger equation.
The resulting BO energy could be furthermore corrected
for the $\Lambda$-doubling, but the numerical value for this correction term
was not clearly identifiable in Ref.~\citenum{KoRy77}.
It would be interesting to see if nonadiabatic corrections can be calculated to
these levels, for example using the recently developed nonadiabatic perturbation theory
by Pachucki and Komasa.\cite{PaKo08,PaKo09}

  The lifetimes of rotational-vibrational levels corresponding to
  the $e\ ^3\Sigma_\mr{u}^+$ state were measured in delayed coincidence experiments,\cite{KiSaAdPaSt99}
  which include both the radiative and predissociative decay channels accessible from these levels.
  In the same work the competition of the two decay channels were investigated
  using ab initio 
  (full configuration interaction electronic wave functions using Gaussian-type orbitals 
  and an accurate adiabatic potential energy curve\cite{KoRy90})
  as well as quantum defect theory with a one-channel approximation.\cite{KiSaAdPaSt99}
  According to these calculations the predissociative lifetimes are of the order
  of $1\ \mu$s and $100$~ns for the $v=0$ and $v=1$ vibrational
  levels, respectively, for the $e\ ^3\Sigma_\mathrm{u}^+$ state with $N=0$.
  The lifetime of the lowest rotational-vibrational levels 
  of the $c\ ^3\Pi_\mathrm{u}^+$ state were calculated 
  using a simple perturbative model, which included the orbit-rotation interaction 
  and used several approximations during the calculations.\cite{ChBh79} 
  In a similar perturbative treatment\cite{CoBr85} the orbit-rotational coupling operator was included
  and accurate BO potential energy curves were used to describe 
  the $b\ ^3\Pi_\mathrm{u}^+$ and $c\ ^3\Pi_\mathrm{u}^+$ states.\cite{KoWo65,KoRy77}
  According to both the lifetime measurements\cite{MeVo78,BrNeLo84} and 
  the calculations\cite{ChBh79,CoBr85} the predissociative lifetime of the lowest-lying 
  rotational-vibrational levels of $c\ ^3\Pi_\mr{u}^+$ are of the order of 1~ns.
  Unfortunately, calculated energy levels were not reported in 
  any of these theoretical work\cite{ChBh79,CoBr85,KiSaAdPaSt99} on 
  the predissociative lifetimes of the $e\ ^3\Sigma_\mr{u}^+$ 
  and $c\ ^3\Pi_\mathrm{u}^+$ states,
  which makes comparison of the results more difficult.
  
  To pinpoint the resonance energies and especially the widths of the rovibronic levels
  within the rigorous pre-BO framework developed in the present work
  further calculations are necessary.
  Nevertheless, the results shown in \ref{tab:resonH2} improve 
  on the best available theoretical values for energy levels 
  published in the literature.\cite{KoRy77,KoRy90}
  Furthermore, our primary goal was also completed, it was
  demonstrated that in pre-BO calculations
  (a) electronically excited rovibronic levels are accessible; and
  (b) there are excited rovibronic levels, which are described as bound states in the BO theory, 
      but appear as resonances in a pre-BO description, i.e.,      
      if the introduction of the BO approximation is completely avoided.

\vspace{0.1cm}
\paragraph{How to improve on the present results?}
First of all, one of the lessons of the present study is that an efficient parameterization
of the basis set is one of the main challenges of the calculation
of rovibronic resonances in pre-BO theory.

We have shown that the random generation of parameters can be improved
using the parameter-transfer approach assuming that there are bound states of comparable
internal structure to the quasi-bound states to be calculated.
In the case of H$_2$
the presented calculations could be improved by the tight optimization of parameter sets
for the lowest-lying bound states with unnatural-parity, $p=(-1)^{N+1}$, 
and by the inclusion of also these parameters in an extended parameter set.

Then, another technically straightforward, but
computationally more demanding option is the optimization of parameters not only
for the lowest-lying states of a symmetry block but for more (or all) vibrational and vibronic excited bound states,
\emph{e.g.} for all (ro)vibrational states of $\X1Sgp$ up to the H(1)+H(1) threshold or
for all the bound (ro)vibronic energy levels
corresponding to the ``$B\ ^1\Sigma_\mr{u}^+$ block'' as well as to the ``$a\ ^3\Sigma_\mr{g}^+$ block''
up to their lowest-lying correlating threshold, H(1)+H(2) (see \ref{fig:orientH2}).

Finally, a more generally applicable solution to the parameterization problem of resonance states
would be the development of a useful and practical application of 
the complex variational principle for resonances.\cite{Moi98}

\clearpage
\begin{table}
  \caption{%
    Assessment of the basis set parameterization:
    the lowest-lying bound-state energies. \\[-0.6cm]
    \label{tab:boundH2}
  }
  \begin{tabular}{@{}c cr cc@{}}
    \cline{1-5} \\[-0.4cm]
    \cline{1-5} \\[-0.3cm]
    \multicolumn{1}{c}{$(N,p,\Sp,\Se)$ $^\mr{a}$} &
    \multicolumn{1}{c}{$E/E_\mr{h}$ $^\mr{b}$} &
    \multicolumn{1}{c}{$\Delta E_\mr{Ref}/\mu E_\mr{h}$ $^\mr{c}$} &
    \multicolumn{1}{c}{Ref.} &
    \multicolumn{1}{l}{Assignment$^\mr{d}$} \\
    \cline{1-5} \\[-0.3cm]
    $(0,+1,0,0)$ &  $-1.164\ 025\ 030$  & $-0.000\ 6$ & $ $\cite{PaKo09}    & $\X1Sgp$ \\
    $(1,-1,1,0)$ &  $-1.163\ 485\ 171$  & $-0.001\ 4$ & $ $\cite{PaKo09}    & $\X1Sgp$ \\
    $(2,+1,0,0)$ &  $-1.162\ 410\ 408$  & $-0.001\ 9$ & $ $\cite{PaKo09}    & $\X1Sgp$ \\
    \cline{1-5} \\[-0.3cm]
    $(0,+1,1,0)$ &  $-0.753\ 027\ 186$  &  $1.383\ 7$ & $ $\cite{WoOrSt06}  & $\B1Sup$ \\
    $(1,-1,0,0)$ &  $-0.752\ 850\ 233$  &  $1.444\ 6$ & $ $\cite{WoOrSt06}  & $\B1Sup$ \\
    $(1,+1,1,0)$ &  $-0.752\ 498\ 022$  &  $1.529\ 1$ & $ $\cite{WoOrSt06}  & $\B1Sup$ \\
    \cline{1-5} \\[-0.3cm]
    $(0,+1,0,1)$ &  $-0.730\ 825\ 193$  & $-0.006\ 9$ & $ $\cite{Wo07}      & $\a3Sgp$ \\
    $(1,-1,1,1)$ &  $-0.730\ 521\ 418$  &  $0.008\ 0$ & $ $\cite{Wo07}      & $\a3Sgp$ \\
    $(2,+1,0,1)$ &  $-0.729\ 916\ 268$  &  $0.047\ 9$ & $ $\cite{Wo07}      & $\a3Sgp$ \\
    \cline{1-5} \\[-0.3cm]
    $(0,+1,1,1)$ & $[-0.999\ 450\ 102]$ $^\mr{e}$ &  $[-5.578]$ & $^\mr{f}$           & $\b3Sup$ \\
    $(1,-1,0,1)$ & $[-0.999\ 445\ 835]$ $^\mr{e}$ &  $[-9.844]$ & $^\mr{f}$           & $\b3Sup$ \\
    $(2,+1,1,1)$ & $[-0.999\ 439\ 670]$ $^\mr{e}$ & $[-16.010]$ & $^\mr{f}$           & $\b3Sup$ \\
    \cline{1-5} \\[-0.4cm]
    \cline{1-5} \\[-0.3cm]
  \end{tabular}
  \begin{flushleft}
    $^\mr{a}$~%
      $N$: total spatial angular momentum quantum number;
      $p:$ parity, $p=(-1)^N$;
      $\Sp$ and $\Se$: total spin quantum numbers for the protons and the electrons, respectively. \\
    $^\mr{b}$~%
      $E$: the energy obtained with the largest parameter set, $\mathcal{P}_L$, used in this study
      corresponding to $15\ 500$ basis functions for each set of quantum numbers
      (see the text for details and the \som\ for the numerical values).
      The proton-electron ratio was $m_\mr{p}/m_\mr{e}=1\,836.152\,672\,47$.\cite{codata06} \\
    $^\mr{c}$~%
      $\Delta E_\mr{Ref}=E_\mr{Ref}-E$ with $E_\mr{Ref}$ being
      the best available non-Born--Oppenheimer theoretical energy value in the literature. \\
    $^\mr{d}$~%
      Born--Oppenheimer electronic state label.
      Each energy level given here can be assigned to the lowest-energy vibrational level of
      the electronic state. \\
    $^\mr{e}$~%
      The lowest-energy eigenvalue of the Hamiltonian obtained for the given set of quantum numbers. \\
    $^\mr{f}$~%
      The non-relativistic energy of two ground-state hydrogen atoms,
      $E(\mr{H}(1)+\mr{H}(1))=-0.999\ 455\ 679\ \mr{E}_\mr{h}$,
      was used as reference. \\
  \end{flushleft}
\end{table}

\begin{table}
  \caption{%
    Identified resonance-state energies and widths, in \Eh, of
    H$_2$ in the $\b3Sup$ block $[p=(-1)^N,S_\mr{p}=(1+p)/2,S_\mr{e}=1]$
    for $N=0,1,$ and $2$.
    \label{tab:resonH2}
  }
  \begin{tabular}{@{} c c@{\ }c ccc @{}}
    \cline{1-6} \\[-0.4cm]
    \cline{1-6} \\[-0.3cm]
    \multicolumn{1}{c}{$(N,p,S_\mr{p},S_\mr{e})$ $^\mr{a}$} &
    \multicolumn{1}{c}{$\mr{Re}(\mathcal{E})$ $^\mr{b}$} &
    \multicolumn{1}{c}{$\Gamma/2$ $^\mr{b}$} &
    \multicolumn{1}{c}{$E_\mr{Ref,exp}$ $^\mr{c}$} &
    \multicolumn{1}{c}{$E_\mr{Ref,theo}$ $^\mr{d}$} &
    \multicolumn{1}{c}{Assignment $^\mr{e}$} \\
    \cline{1-6} \\[-0.3cm]
    $(0,+1,1,1)$ & $[-0.999\ 450\ 1]$ $^\mr{f}$ & & & $[-0.999\ 455\ 7]$ & H(1)+H(1) continuum \\
    & $[...]$ & & & & \\
    $(0,+1,1,1)$ & $-0.677\ 947\ 1$ & $1\cdot 10^{-7}$ & $-0.677\ 946\ 1$ & $-0.677\ 942\ 7$ $ $\cite{KoRy90} & $e\ ^3\Sigma_\mr{u}^+, R=0, v=0$ \\
    $(0,+1,1,1)$ & $-0.668\ 549\ 3$ & $9\cdot 10^{-7}$ & $-0.668\ 547\ 8$ & $-0.668\ 541\ 0$ $ $\cite{KoRy90} & $e\ ^3\Sigma_\mr{u}^+, R=0, v=1$ \\
    \cline{1-6} \\[-0.3cm]
    $(1,-1,0,1)$ & $[-0.999\ 445\ 8]$ $^\mr{f}$ & & & $[-0.999\ 455\ 7]$ & H(1)+H(1) continuum \\
    & $[...]$ & & & & \\
    $(1,-1,0,1)$ & $-0.731\ 434\ 0$ & $5\cdot 10^{-7}$ & $-0.731\ 438\ 8$ & $-0.731\ 469\ 1$ $ $\cite{KoRy77} & $c\ ^3\Pi_\mr{u}^+, R=0, v=0$ \\
    $(1,-1,0,1)$ & $-0.720\ 72$ & $2\cdot 10^{-7}$ & $-0.720\ 782\ 6$ &                  & $c\ ^3\Pi_\mr{u}^+, R=0, v=1$ \\
    & $[...]$ & & & & \\
    $(1,-1,0,1)$ & $-0.677\ 705\ 5$ & $2\cdot 10^{-7}$ & $-0.677\ 704\ 1$ & $-0.677\ 698\ 2$ $ $\cite{KoRy90} & $e\ ^3\Sigma_\mr{u}^+, R=1, v=0$ \\
   $(1,-1,0,1)$ & $-0.668\ 32$ & $1\cdot 10^{-6}$ & $-0.668\ 319\ 7$ & $-0.668\ 309\ 8$ $ $\cite{KoRy90} & $e\ ^3\Sigma_\mr{u}^+, R=1, v=1$ \\
    \cline{1-6} \\[-0.3cm]
    $(2,+1,1,1)$ & $[-0.999\ 439\ 7]$ $^\mr{f}$ & & & $[-0.999\ 455\ 7]$ & H(1)+H(1) continuum \\
    & $[...]$ & & & & \\
    $(2,+1,1,1)$ & $-0.730\ 888\ 2$ & $9\cdot 10^{-7}$ & $-0.730\ 888\ 7$ &                  & $c\ ^3\Pi_\mr{u}^+, R=1, v=0$ \\
    $(2,+1,1,1)$ & $-0.720\ 219\ 0$ & $<2\cdot 10^{-7}$& $-0.720\ 258\ 0$ &                  & $c\ ^3\Pi_\mr{u}^+, R=1, v=1$ \\
    & $[...]$ & & & & \\
    $(2,+1,1,1)$ & $-0.677\ 222\ 9$ & $2\cdot 10^{-8}$ & $-0.677\ 222\ 2$ &                  & $e\ ^3\Sigma_\mr{u}^+, R=2, v=0$ \\
    $(2,+1,1,1)$ & $-0.667\ 863\ 2$ & $7\cdot 10^{-7}$ & $-0.667\ 865\ 3$ &                  & $e\ ^3\Sigma_\mr{u}^+, R=2, v=1$ \\
    \cline{1-6} \\[-0.4cm]
    \cline{1-6} \\[-0.3cm]
  \end{tabular}
  \begin{flushleft}
    ~\\[-0.5cm]
    $^\mr{a}$~%
      $N$: total spatial angular momentum quantum number;
      $p:$ parity, $p=(-1)^N$;
      $\Sp$ and $\Se$: total spin quantum numbers for the protons and electrons, respectively. \\
    $^\mr{b}$ %
      $\mr{Re}(\mathcal{E})$ and $\Gamma$:
      resonance energy and width with $\Gamma/2=-\mr{Im}(\mathcal{E})$ calculated in this work.
      The largest basis set contained $15\ 500$ basis functions for each set of quantum numbers.
      The proton-electron ratio was $m_\mr{p}/m_\mr{e}=1\,836.152\,672\,47$.\cite{codata06} \\
    $^\mr{c}$~%
      $E_\mr{Ref,exp}$
      experimental reference value, in \Eh, derived as $E_\mr{exp}=E_0+T_\mr{exp}$ with
      the ground-state energy ($X\ ^1\Sigma_\mr{g}^+,N=0,v=0$) $E_0=-1.164\ 025\ 030$~\Eh.
      All $T_\mr{exp}$ values were obtained by correcting the experimental term values of Dieke\cite{Di58},
      with $-0.000\ 681\ 7\ \mr{E}_\mr{h} = -149.63\ \mr{cm}^{-1}$
      ($1\ \mr{E}_\mr{h}=219\ 474.631\ 4\ \mr{cm}^{-1}$),
      since all triplet term values were too high.\cite{MiFr74,KoRy90} \\
    $^\mr{d}$~%
      $E_\mr{Ref,theo}$: the best available theoretical reference energy values, in \Eh,
      corresponding to accurate adiabatic calculations for the $e\ ^3\Sigma_\mr{u}^+$ levels\cite{KoRy90}
      and to accurate Born--Oppenheimer calculations for the $c\ ^3\Pi_\mr{u}^+$ levels.\cite{KoRy77}
      The non-relativistic energy of two ground-state hydrogen atoms
      is given in square brackets. \\
    $^\mr{e}$~%
      Born--Oppenheimer electronic- and vibrational-state labels.
      The (approximate) rotational angular momentum quantum number, $R$, is also given. \\
    $^\mr{f}$~%
      The lowest-energy eigenvalue of the real Hamiltonian obtained with the
      largest parameter set and with the given quantum numbers.
  \end{flushleft}
\end{table}

%
%
\clearpage
\section{Summary and outlook\label{ch:sum}}
The present work was devoted to the calculation of rotational-vibrational energy levels
corresponding to electronically excited states, which are bound within
the Born--Oppenheimer (BO) approximation
but appear as resonances in a pre-Born--Oppenheimer (pre-BO) quantum mechanical description.

In order to calculate resonance energies and widths, corresponding to predissociative lifetimes,
the pre-BO variational approach and computer program of Ref.~\citenum{MaRe12}
was extended with the complex coordinate rotation method (CCRM).
Similarly to the bound-state calculations, the wave function was written as a linear combination
of basis functions which have the non-relativistic quantum numbers
(total spatial---rotational plus orbital---angular momentum quantum number,
parity, and total spin quantum number for each particle type).
The basis functions were constructed using
explicitly correlated Gaussian functions and the global vector representation.

This pre-BO-resonance approach was first used for
the three- and four-particle positronium complexes,
Ps$^-=\lbrace\mr{e}^-,\mr{e}^-,\mr{e}^+\rbrace$ and
Ps$_2=\lbrace\mr{e}^-,\mr{e}^-,\mr{e}^+,\mr{e}^+\rbrace$, respectively.
These applications allowed us to test the implementation and gain experience in the identification of
resonance parameters. For the dipositronium, Ps$_2$, we managed to improve on some of the best
available results reported in the literature.

Then, the developed methodology and technology was employed for the four-particle molecule, H$_2$.
First, the rovibronic states known in the literature
were collected and considered which were accessible in our calculations with
various sets of (exact) quantum numbers of the non-relativistic theory.
The experimental and theoretical energy values of the literature
were also used for the assignment of our calculated energy levels with
the common BO terminology of electronic and vibrational state labels.

As to the computational part, we had to find a useful parameterization strategy for the basis functions.
Since the bound-state parameter-optimization approach relied on the energy minimization condition and
the (real) variational principle, it was not directly applicable for making the CCRM calculations more efficient.
A simple and practical solution to the parameterization problem was the parameter-transfer approach,
the basis functions used to describe low-energy resonances of some symmetry were parameterized
with optimized parameters of (high-energy) bound states.
As a result, a large parameter set was constructed, which was compiled from
parameters optimized for different symmetry blocks. The parameterization of basis functions,
which have mathematical forms defined by the exact quantum numbers, with this extended parameter set
immediately lead to an improvement for the best available energy values available in the literature
for the lowest-lying rotational states assigned to the $\B1Sup$ and $\a3Sgp$ electronic states.

Then, using this extended parameter set low-energy rovibronic resonances became accessible
beyond the $\b3Sup$ repulsive electronic state, embedded in the H(1)+H(1) continuum.
Based on these calculations, resonance energies were evaluated and resonance widths were estimated for
the lowest-lying rotational-vibrational levels
of the electronically excited $e\ ^3\Sigma_\mr{u}^+$ and $c\ ^3\Pi_\mr{u}^+$ states.
We note here that the coupling of the rotational and orbital angular momenta was automatically
included in our computational approach by specifying only the total spatial angular momentum quantum number, $N$.
Although the presented results improve on the best available (BO and adiabatic) calculations in the literature for these states,
to pinpoint the resonance energies and especially the widths more extensive calculations are necessary.

As to further improvements, %
the major present technical difficulty is the efficient parameterization
of the basis set for resonance states. A generally applicable solution to this problem would be
a useful application and implementation of a complex analogue for the real variational principle.
In the lack of such a general solution, optimization of large parameter sets for bound (excited) states together
with the parameter-transfer strategy might be appropriate for the calculation of a larger number and/or
more accurate rovibronic resonances of H$_2$.
In addition to the improvement of the parameterization strategy,
a generally applicable analysis tool would also be desirable
which provides an assignment for the pre-BO wave function
with the common BO electronic- and vibrational-state labels where
such an assignment is possible.

\acknowledgement
E.M. is thankful to Prof. Attila G. Cs\'asz\'ar and Prof. Markus Reiher for the continuous encouragement
and also thanks Benjamin Simmen for discussions.
The financial support of the Hungarian Scientific Research Fund (OTKA, NK83583) 
an ERA-Chemistry grant is gratefully acknowledged.
The computing facilities of HPC-Debrecen (NIIFI) were used during this work.

\paragraph{Supporting Information Available} 
This information is available free of charge via the Internet at http://pubs.acs.org .



\begin{mcitethebibliography}{91}
\bibitem[spe()]{spectracc}
The term ``spectroscopic accuracy'' is not uniquely defined but it is usually
  used to refer to calculations providing vibrational transition wave numbers
  with a certainty of at least 1~\cm\ ($\approx 4.6\ \mu$E$_\mr{h}$) and even
  better accuracy for calculated rotational transitions.\relax
\mciteBstWouldAddEndPunctfalse
\mciteSetBstMidEndSepPunct{\mcitedefaultmidpunct}
{}{\mcitedefaultseppunct}\relax
\EndOfBibitem
\end{mcitethebibliography}

\providecommand*\mcitethebibliography{\thebibliography}
\csname @ifundefined\endcsname{endmcitethebibliography}
  {\let\endmcitethebibliography\endthebibliography}{}


\end{document}